\documentclass{article}

\usepackage{arxiv}

\usepackage[utf8]{inputenc}
\usepackage[T1]{fontenc}
\usepackage{CJKutf8}
\usepackage{hyperref}       
\usepackage{url}            
\usepackage{booktabs}       
\usepackage{amsfonts}       
\usepackage{nicefrac}       
\usepackage{microtype}      
\usepackage{amsmath}
\usepackage{lipsum}
\usepackage{float}
\usepackage{tabularx}
\usepackage{graphicx}
\graphicspath{ {./images/} }

\title{JaColBERTv2.5: Optimising Multi-Vector Retrievers to Create State-of-the-Art Japanese Retrievers with Constrained Resources}

\author{
 Benjamin Clavié \\
  Answer.AI \\
  \texttt{bc@answer.ai} \\
}

\begin{document}
\maketitle
\begin{abstract}
Neural Information Retrieval has advanced rapidly in high-resource languages, but progress in lower-resource ones such as Japanese has been hindered by data scarcity, among other challenges. Consequently, multilingual models have dominated Japanese retrieval, despite their computational inefficiencies and inability to capture linguistic nuances. While recent multi-vector monolingual models like JaColBERT have narrowed this gap, they still lag behind multilingual methods in large-scale evaluations.
This work addresses the suboptimal training methods of multi-vector retrievers in lower-resource setting, focusing on Japanese. We systematically evaluate and improve key aspects of the inference and training settings of JaColBERT, and more broadly, multi-vector models. We further enhance performance through a novel checkpoint merging step, showcasing it to be an effective way of combining the benefits of fine-tuning with the generalization capabilities of the original checkpoint.
Building on our analysis, we introduce a novel training recipe, resulting in the JaColBERTv2.5 model. 
JaColBERTv2.5, with only 110 million parameters and trained in under 15 hours on 4 A100 GPUs, significantly outperforms all existing methods across all common benchmarks, reaching an average score of 0.754, significantly above the previous best of 0.720.
To support future research, we make our final models, intermediate checkpoints and all data used publicly available.
\end{abstract}

\section{Introduction}

Text-based Information Retrieval (\textit{IR})~\cite{kobayashi,ir99} is a Natural Language Processing application that enables users to retrieve relevant documents for a given query~\cite{manning}. Historically, the field has been dominated by lexical-matching methods, where retrieval is performed by directly matching query terms with document content~\cite{lexicalmatch}, with both undergoing various normalization steps to facilitate this process~\cite{stemming}. Okapi BM25~\cite{bm25}, the most successful lexical-matching method, remains widely used today, either independently or in conjunction with more modern approaches~\cite{beir}.

Despite its inability to capture semantic information, BM25 is still considered a robust baseline that can be challenging to surpass, even with significantly more computationally intensive approaches~\cite{bm25strong}. However, in recent years, there have been large advances in the development Neural IR approaches for the English language, with numerous models now consistently strongly outperforming BM25~\cite{mteb}.

These improvements have been attributed to two factors: the first is the development of much more powerful neural architectures, such as the transformers architecture~\cite{transformers}, and the pre-trained models based on it such as BERT~\cite{bert}. The second factor is the release of large-scale retrieval training datasets, with Microsoft's MS-MARCO~\cite{msmarco} being widely credited as one of the most important factors to the rapid development of English Neural IR~\cite{mrtydi}. Moreover, there also exists a large range of high quality, domain-specific datasets on which Neural retrievers can be further trained, providing English models with even stronger generalisation capabilities~\cite{trec,beir}.

Outside of the English language, and some other very high-resources languages such as Mandarin Chinese~\cite{cpack}, progress in Neural IR has been considerably slower and has yielded more modest results. In most mid and lower-resource languages, as is the case in Japanese, retrieval benchmarks have largely been dominated by multilingual models~\cite{me5,m3}, taking advantage of the vast quantities of English training data to bolster their performances on all languages. 

However, these models come with multiple trade-offs. Noticeably, their retrieval performance ceiling appears to be lower than monolingual models, as, in the case of English, monolingual retrievers trained on the same amount of English consistently show stronger performances than their multilingual counterpart~\cite{e5,bge}. Their relative compute costs are also considerably higher, both at training time, as they require several billion query-document pairs, and inference time, with a similarly performing multilingual retriever having 3-to-5 times as many parameters as monolingual variants~\cite{jacolbert,m3,xlmr}. Finally, multilingual models have been empirically shown to miss out on cultural understanding in some settings~\cite{singaporecultural}, which represents an ethical issue as well as a potential performance one in certain settings.

In the case of Japanese, the performance gap has been particularly wide, with monolingual neural models often representing a 40\% average performance decrease from multilingual ones, with even starker degradation on larger-scale benchmarks~\cite{jacolbert}.

All existing mono-lingual Japanese retrieval models have followed the most common approach to Neural IR, which consists in the use of single-vector dense representations, where a document is represented as a single averaged vector~\cite{dpr}. However, multi-vector retrievers, where documents are represented by multiple vectors, one for each token it contains, have shown remarkable training efficiency and out-of-domain generalisation since their introduction with the ColBERT model~\cite{colbert}, and its further refinement in ColBERTv2~\cite{colbertv2}.  The promising results of ColBERT models in English and their strong generalisation ability while training on a single domain, appeared to suggest the potential suitability of this approach in lower-resources settings, such as the Japanese language.

Following these findings, a recent study has introduced Japanese ColBERT variants, the JaColBERT and JaColBERTv2 models~\cite{jacolbert}, respectively following the ColBERT and ColBERTv2 training recipes. These models, while trained exclusively on lower-quality Japanese training data, have demonstrated considerable performance improvements. They have vastly outperformed existing monolingual retrievers and reached competitive results with the best multilingual ones, albeit still falling short on large scale benchmarks containing millions of documents.

In spite of this, while there has been considerable improvements in understanding how to best train both general NLP~\cite{adamw,wsd,schedulefree,gpt4,llama2} and retrieval models~\cite{colberter,gravedistill,m3,linbaselines}, the training of ColBERT models has followed the standard ColBERTv2 recipe with some changes, but no systematic overhaul or in-depth modifications~\cite{colbertx,jacolbert,colbertxm}.

\subsection{Contributions}

In this study, we present the hypothesis that it is possible to outperform all existing retrieval methods, including the best multilingual models, in Japanese with an optimized multi-vector retriever training recipe.

To achieve this aim, we systematically evaluate potential improvements via many small-scale ablation studies, seeking to increase both training efficiency and downstream performance.

We begin by evaluating inference-time setting, and demonstrate that dynamic query-length is strictly superior to fixed-length querying.

The impact of using various teacher models for knowledge distillation is then systematically evaluated, as well as the performance changes when ensembling various teachers, as is oftentimes the best option in English~\cite{marginmse}. We show that this does not hold true in a Japanese setting, and generate teacher scores using the best-performing model in our ablation studies, BGE-M3~\cite{m3}.

Our subsequent experiments then demonstrate that conventionally used training settings for ColBERT models are either computationally inefficient, place unnecessary constraint on data gathering for no performance improvement, or are simply sub-optimal in terms of retrieval performance. Based on our results, we propose an improved training setting.

We propose an additional two final steps to the training process. The first one is a post-training step, where we leverage data with higher-quality Japanese to fine-tune our final model checkpoint, in the hopes of improving its results on common Japanese retrieval tasks. We then introduce a final checkpoint averaging step, where the models resulting from this post-training steps are averaged with checkpoints from the pre-training phase, to create a model that retains the performance gains on tasks which are in-domain for the post-trained model, without losing any performance on out-of-domain tasks, further increasing the generalisation potential of our model.

Our resulting model, JaColBERTv2.5, is the best performing model on all evaluation datasets, reaching an average score of 0.752, representing a 4.5\% improvement over JaColBERTv2 (0.720), a 60\% performance improvement on the best-performing monolingual single-vector retriever, GLuCoSE (0.470) and a 5.32\% improvement over BGE-M3 (0.714), the strongest multilingual model. These results hold true for MIRACL~\cite{miracl}, a large-scale retrieval benchmark on which previous versions of JaColBERT trailed significantly behind BGE-M3, but JaColBERTv2.5 reaches a 6.87\% performance improvement over it.

Our model prior to the post-training step, JaColBERTv2.4, also outperforms all other existing approaches, even while being fully out-of-domain on every evaluation dataset.

We achieve these results with a constrained compute budget, where the compute used for our final model, teacher score generation included, cannot meaningfully exceed that of JaColBERTv2, to confirm that downstream performance comes from our improved recipe rather than from increased computational budget.

Moreover, we obtain these results with a simplified training recipe, which fully discards the "positive" and "negative" labels assigned to each document, focusing entirely on the relative distribution of teacher scores instead. These results can help support future research in the area of data curation and hard-negative mining, and help simplifying both processes.

We make all the resources in this paper publicly available to support further research. This includes all of our training data, with teacher scores, for both the ablation and final training runs\footnote{\url{https://huggingface.co/datasets/answerdotai/MMarco-japanese-32-scored-triplets}},
the JaColBERTv2.5\footnote{\url{https://huggingface.co/answerdotai/JaColBERTv2.5}} 
and JaColBERTv2.4 models\footnote{\url{https://huggingface.co/answerdotai/JaColBERTv2.4}},
as well as all intermediate model checkpoints generated during training\footnote{\url{https://huggingface.co/collections/bclavie/jacolbertv25-checkpoints-66a37d8da6b0d4d69c14f9c3}}.

Finally, while our study focused entirely on the Japanese language, we believe that our method can be directly applicable to any other language with at least moderate data volumes and lead to similar performance improvements, as even our 320,000 triplets vastly outperformed previous monolingual approaches.

\section{Background}

In this section, we will provide a brief overview of the mechanisms used by multi-vector retrieval models such as ColBERT and JaColBERT, on which our work builds.

\subsection{ColBERT}
\label{sec:colbertintro}

Multi-vector retrievers, also sometimes called late-interaction models, were introduced and popularised by the Contextualized Late Interaction over BERT, or ColBERT, model architecture~\cite{colbert}.

The ColBERT architecture relies on a simple intuition, which is that the dominant way of creating document representation, by creating a single vector for each document, causes too much information loss. While this information loss can be mitigated when evaluated on in-domain tasks, it can result in consistently poorer out-of-domain generalisation. To solve this problem, multi-vector retrievers do not create a single large vector to represent a document, but multiple smaller ones with each individual vector representing a single token.

\textbf{Query Augmentation} Identifying the fact that retrieval is by nature a very asymmetrical task, with queries often being short, a query-augmentation mechanism is also introduced. Rather than padding inputs to the maximum length with padding tokens, it leverages the masked-language-modeling objective of its base models~\cite{bert}. To do so, it pads all queries with [MASK] tokens, which are then attended to by the model. These mask tokens have been heavily studied in subsequent work, and appear to provide useful term-weighting and amplify semantic information, resulting in improved retrieval performance.~\cite{mask_analysis,mask2}

\textbf{MaxSim} In order to perform scoring, the ColBERT authors introduce a new scoring mechanism, which they dub MaxSim, for Maximum Similarity. In this setting, the score of a [query, document] pair is obtained via the following formula, where E represents all the embeddings representing a given document or query:

\begin{equation}
    Score_{query,document} := \sum_{i \in [|E_{query}|]} \max_{j \in [|E_{document}|]} E_{{query}_i} * E_{{document}_j}^{T}    
\end{equation}
    
Effectively, for a given query token, its cosine similarity with every document token is computed, and the highest similarity is kept. This process is repeated over all query tokens, and the final [query, document] score is represented as the sum of all those maximum similarities per query token, hence the name "MaxSim".

\textbf{ColBERT Initial Performance} This approach appears to consistently result in better out-of-domain generalisation, even when trained on considerably smaller data volumes than competitive single-vector approaches~\cite{colbert,colberter}. However, its performance on in-domain tasks remained lower than that of single-vector retrievers, while requiring an order of magnitude more memory and storage space to index the same document volume due to the need of storing significantly more vectors. \\

\subsection{ColBERTv2}
A subsequently improved version of ColBERT, named ColBERTv2, seeks to address these two issues~\cite{colbertv2}. This second version overhauls many parts of the original process, using a more modern training recipe, albeit without a clear evaluation of the impact of each component. Most notably on the training side, it introduces both in-batch negatives~\cite{ibneg} and knowledge distillation~\cite{distillation}. To help alleviate the storage issue, it introduces a novel indexing approach, allowing for a 6-fold index-size reduction by clustering the individual token representations and offloading most of the stored information to the cluster centroids before compressing the vectors to just 2 bits. This method successfully brings the storage and memory requirements of ColBERTv2 down to the same order of magnitude as that of single-vector models, though still noticeably higher, while reaching even stronger results on out-of-domain datasets. Further work has shown that this approach also addresses the issue of weaker in-domain performance, with fine-tuned versions of the model being able to outperform all other approaches on multiple benchmarks~\cite{udapdr}.

ColBERT and ColBERTv2 have garnered a lot of attention, with various studies attempting to better understand and improve its various mechanisms, such as the various effects of [MASK]-based query augmentation~\cite{mask2,mask_analysis}, the impact of introducing full-word rather than token-level representations~\cite{colberter}, potential improvements to its scoring approach~\cite{xtr}, or to the mechanisms around its optimised indexing approach~\cite{plaid,plaidrepro}.\\

\subsection{JaColBERT}

In Japanese, both of these approaches have been reproduced, with even greater success than their English equivalents~\cite{jacolbert}. JaColBERTv1, following the training recipe of the original ColBERT model, became the then-strongest monolingual Japanese retriever. However, it fell short of the strongest multilingual models on multiple benchmarks, with the most notable performance gap being found on large-scale retrieval tasks. Subsequently, JaColBERTv2, trained following the ColBERTv2 recipe, helped address these issues. JaColBERTv2, at the time of this work, is the strongest out-of-domain retrievers on all existing Japanese benchmarks. However, on MIRACL~\cite{miracl}, a large-scale retrieval dataset which was used to train most multilingual retrievers and on which they are therefore in-domain, it still noticeably lags in performance.

\section{Experiments}

In this section, we will present the various steps of our experimental process. As this study focuses on systematically evaluating various approaches to using and training multi-vector models, we will conduct short training runs, also called \textit{ablations}, on small data scales, to identify the best possible setting. We believe this sort of small-scale training to identify optimal model settings is well-suited to helping us refine optimal training for constrained resources settings, as it has proved to be a particularly strong indicator of full-sized model performance. In recent months, it has notably become the preferred way of predicting model behaviour for large language models~\cite{gpt4,predictscaling,llama2}. 

In each section, we will discuss the rationale for our proposed settings and, when relevant, the results and learnings of the relevant ablations.

Firstly, we will define our hardware constraints in Section~\ref{sec:hardware}, before discussing our choice of training data is Section~\ref{sec:trainingdata}. We then present an overview of our the baselines our models will be evaluated against in Section~\ref{sec:evalbaselines} and of our chosen valuation benchmarks in Section~\ref{sec:eval}.

In Section~\ref{sec:querylen}, we will explore different approaches to defining the query length on ColBERT's query-augmentation mechanism.

We will then evaluate the impact of various alterations of the model's training settings in Section~\ref{sec:trainsettings}, through a series of small-scale training runs.

Our next area of focus in Section~\ref{sec:teacher} is systematically studying the use of various teacher models for knowledge distillation, in the context of Japanese-language retrieval. We will explore the impact of using a variety of models as teachers, as well as various ensembling methods.

Finally, we propose two ways of improving our final model, after the original pre-training has concluded. In Section~\ref{sec:posttrain}, we describe a post-training phase, where our model will be fine-tuned on smaller-scale, considerably higher-quality data. In Section~\ref{sec:averaging}, we discuss the introduction of checkpoint averaging~\cite{polyak3} as a method to ensure that the final model remains strong across the board, and mitigating potential performance losses on out-of-domain tasks during post-training.

For all training runs, both small and large scale, we follow JaColBERTv2 in initializing our models off the original JaColBERT checkpoint, which has been trained on a total of 8 million [query, positive\_document, negative\_document] training triplets~\cite{jacolbert}.

\subsection{Hardware Constraints}
\label{sec:hardware}

All experiments conducted in this study are done under a compute constraint, in order to highlight that our final model's performance is a consequence of our improvements, rather than due to substantially increased compute.

JaColBERTv2 was trained for 28.5 hours on 8 NVidia A100 GPUs~\cite{jacolbert}, representing a total training budget of 228 A100 hours. All teacher scores used by JaColBERTv2 were re-used from the original ColBERTv2 teacher scores~\cite{colbertv2}, and therefore came at no additional compute cost.

Based on this, we constrain our total training compute budget to the same 228 hours, plus or minus 5\%, resulting in a final upper bound budget of \textbf{239 A100 hours}. We include all time spent training each ablation model, generating teacher scores and training the final JaColBERTv2.5 model under this budget.

\subsection{Training Data}
\label{sec:trainingdata}

For both the final model training and ablation runs, we follow existing practices~\cite{jacolbert,me5} and train our model using the Japanese split of MMarco~\cite{mmarco}. MMarco is a machine-translated of MS Marco~\cite{msmarco}, a large English Information Retrieval (IR) dataset which has widely been credited as unlocking vast advances in English IR, thanks in large part to its scale and wide variety of queries. For a long time, no equivalent dataset existed in non-English languages, and efforts to create ones were largely unsuccessful due to the cost of such an endeavour. MMarco was introduced to bridge the gap between English and other languages, by providing a fully machine-translated version of MS Marco in 14 languages, and empirically showcasing that, while the resulting dataset produced poorer results than in English, it still contained useful signal on a scale usually not available in these languages.

Retrievals models are generally trained on triplets. These triplets can either be be standard triplets, as with ColBERTv1~\cite{colbert} and JaColBERTv1 described above, where each triplet contains a single query, a single positive document, and a single negative document, or n-way triplets. n can be any number, and represents the total number of documents passed to the model: in the case of 16-way triplets, the model would be presented with a query, and 16 documents, rather than just 2 in the standard setting. Doing so allows us to more efficiently use \textit{knowledge distillation}~\cite{distillation}, where the model learns from teacher scores, generally generated by strong cross-encoder models, and attempts to emulate them or their distribution~\cite{rocketqa}.

Our models are trained using 32-way triplets with knowledge distillation. This means that, for every single query, the model is given 32 documents per triplet, as well as teacher scores for every [query, document] pair. The goal of the model's training is to attempt to learn to reproduce the provided scores, through a knowledge distillation loss function which we explore further in Section~\ref{sec:loss}.

We use a downsample of the set of triplets used to train the original English ColBERTv2 model~\cite{colbertv2}. We downsample the triplets in two ways: firstly, in order to meet the compute constraints of this work, we randomly sample 3,200,000 triplets out of the 19,000,000 originally provided.
Secondly, as ColBERTv2 was trained with 64-way triplets, we randomly sample 31 negative documents from the original 63 in every individual triplet. We choose to train on 3,200,000 triplets, which represents 40\% of the 8,000,000 triplets JaColBERTv2 was trained on, in order to respect our compute constraints and allocate sufficient compute to generating teacher scores.

As MMarco is a direct translation of MS Marco, it is possible to reuse the ColBERTv2 triplets with no further modifications. We use the teacher scores provided by the ColBERTv2 authors as our baseline teacher scores for all of our experiments, and will extensively cover the effect of different teachers in Section~\ref{sec:teacher}. 

In recent years, higher quality datasets such as Mr.TyDi~\cite{mrtydi} and its improved version, MIRACL~\cite{miracl}, have been introduced. However, both of them contain noticeably fewer queries and relevance labels per language than MMarco, and are more commonly used to train multi-lingual models~\cite{m3,me5}, while best-performing monolingual retrievers largely continue to pre-train on MMarco~\cite{decouvrir}. However, we believe that MIRACL could constitute a particularly suitable post-training dataset, which we will further explore in Section~\ref{sec:posttrain}.

\textbf{Ablation Training Data} For all ablation training runs we conduct as part of our experiment, we use a further downsampled version of our final training set. We create this set by sampling the first 10\% triplets of the full training set, resulting in 320,000 training triplets, which represents 4\% of the original JaColBERTv2 training data. Following previous work~\cite{gpt4,predictscaling}, we believe that this data volume is sufficient to show trends which will scale to the final training run.

\subsection{Baselines}
\label{sec:evalbaselines}

Our final models are evaluated against a large range of representative baselines, including the current best-performing retrievers. To do so, we evaluate our model against BGE-M3~\cite{m3}, the current best-performing multilingual embedding model. BGE-M3 is a multi-output model: it is capable of producing single-vector dense representations, but is also able to output sparse and computationally heavy multi-vector representations to act as a "self-reranker". As a result, we report results from BGE-M3 in two settings: \textit{dense}, using only its single-vector retriever output, and \textit{all}, leveraging all three forms of outputs in the way recommended by its original authors. BGE-M3's model size is roughly 5.11x that of JaColBERT.

Results for the multilingual-E5 (mE5) family of models~\cite{me5} are also presented in all three existing model sizes, \textit{small} (\~JaColBERT-sized), \textit{base}(\~2.5x JaColBERT) and \textit{large} (\~5x JaColBERT). The mE5 family is one of the most widely used model family for Japanese retrieval, and has consistently shown strong results on benchmarks~\cite{jacolbert}.

We also report results for the best-performing single-vector retrievers in Japanese, GLuCoSE\footnote{\url{https://huggingface.co/pkshatech/GLuCoSE-base-ja}}, an embedding model based on LUKE~\cite{luke}, as well as Nagoya University's \textsc{sup-simcse} family of models~\cite{nagoya}, in both \textit{base} and \textit{large} sizes.

Finally, we also report results for JaColBERTv1 and JaColBERTv2, the two previous best multi-vector retriever models for Japanese, respectively trained following the ColBERTv1~\cite{colbert} and ColBERTv2~\cite{colbertv2} training recipes.


\subsection{Evaluation Data}
\label{sec:eval}

We define two evaluation sets: one used for final evaluation, described in Section~\ref{sec:evaldata} and a smaller, quicker-to-run one, that will be used for the various experiments we are planning to run to find the optimal training setting presented in Section~\ref{sec:devdata}. All metric calculations are performed using the \textit{ranx} evaluation toolkit~\cite{ranx}.

\subsubsection{Final Evaluation Data}
\label{sec:evaldata}

For the final evaluation, five commonly used evaluation datasets will be used to cover the model's performance in a variety of settings. For each dataset, we choose a main evaluation metric in line with previous work in order to provide clear comparisons. However, detailed evaluation results for our models will also be reported.

\begin{table}[H]
\centering
\begin{tabular}{llccl}
                            & Type                  & \# Queries & \# Documents & Task Setting          \\ \hline
\multicolumn{1}{l|}{MIRACL} & General domain QA     & 860        & 6,953,614    & Large-Scale Retrieval \\
\multicolumn{1}{l|}{JSQuAD} & General domain QA     & 4420       & 1145         & Small-Scale Retrieval \\
\multicolumn{1}{l|}{JQaRA}  & Trivia QA             & 1667       & 100          & Reranking             \\
\multicolumn{1}{l|}{JaCWIR} & Synthetic Web QA      & 5000       & 100          & Reranking             \\
\multicolumn{1}{l|}{ESCI}   & Amazon Product Search & 4206       & 149,999      & Large-Scale Retrieval
\end{tabular}
\caption{Brief overview of the key information about the datasets used for evaluation. We consider as "Large-Scale Retrieval" any task where more than 100,000 documents are considered at once. For Reranking tasks, the number of documents is the number of documents to rerank.}
\label{tab:evaldatasets}
\end{table}

\textbf{JSQuAD} is a QA dataset introduced in the JGLUE evaluation set~\cite{jsquad}, inspired by the English SQuAD~\cite{squad}. We evaluate JSQuAD in the same setting as in previous studies~\cite{jacolbert}, using Nouu.me's evaluation benchmark, where the dataset is reduced to 1600 passages, and the model's goal is to extract the relevant passage for each query in its top results. The metric we report for this dataset, following previous work, is Recall\@3. This task is the easiest amongst the evaluation set.

\textbf{MIRACL}~\cite{miracl} is a large-scale multilingual evaluation benchmarks. We use its Japanese subsplit, which is composed of over 6 million documents, extracted from Wikipedia, and contains human-created relevance judgements for 860 queries over this corpus. We choose to use exclusively MIRACL, rather than both MIRACL and Mr.TyDi~\cite{tydi}, another large-scale multilingual information retrieval dataset, as MIRACL is a refinement of Mr.TyDi, with additional judgements added and dubious labels removed. The main metric reported for this dataset is NDCG@10. It has been noted in the past that MIRACL contains "holes": that is, the positive judgements are not thorough, and the data contains many false negatives\footnote{While there is no formal citation for this claim, it can deduced from the annotation process used for MIRACL and has been frequently noted, notably as part of Cohere's work on multilingual embeddings:
 \url{https://huggingface.co/datasets/Cohere/miracl-en-queries-22-12}}. However, it remains the only large-scale evaluation benchmark for most languages it covers, including Japanese, and is one of the most commonly used non-English IR benchmark~\cite{m3,me5,jacolbert,colbertxm}.

\textbf{JQaRA}~\cite{jqara} is another dataset built from a QA dataset commonly used for Japanese QA evaluation, JAQKET~\cite{JAQKET}. The aim is, similarly to SQuAD, to find a document containing the answer to a given query, over 1667 queries. The dataset was constructed via a mix of LLM usage, before going through human validation to ensure all negatives are negatives and all questions have at least one real positive passage. The task is presented as a hard reranking task: for each query, we are provided with 100 documents, with one or more of them containing the information necessary to answer the query and all other documents containing very adjacent information which does not directly address the query. These adjacent documents are called "hard negatives", as they're purposefully designed to be hard to differentiate from positive examples. The main evaluation for this task is NDCG@10. All evaluations on JQaRa are conduced with the official evaluation code provided by the dataset author~\cite{jqara}.

\textbf{JaCWIR}~\cite{jacwir} is a medium-scale (500,000 documents) retrieval dataset, using a large variety of web-scrapped documents. It is an entirely auto-generated dataset, where GPT-3.5~\cite{gpt35} was asked to produce queries for which a given document would be relevant. We also use it as a reranking task, similarly to how it was introduced. For each of its 5,000 queries, the model must attempt to identify the relevant document among 99 hard negatives. The main evaluation metric for this task is NDCG@10. All evaluations on JaCWIR are conduced with the official evaluation code provided by the dataset author~\cite{jacwir}.

\textbf{ESCI}~\cite{ESCI} is an addition to the JaColBERT evaluation set, which was not used in previous work. It is a \textit{"Shopping Queries Data Set"} 
dataset provided by Amazon Science as part of the KDD Cup 2022. The goal of this dataset is to evaluate a model's ability to match very short (1 to 5 tokens) queries with the textual description of relevant products. We use ESCI as a retrieval task, similarly to one of the settings it is available in in the Japanese Massive Text Embeddings Benchmark (JMTEB)\footnote{\url{https://huggingface.co/datasets/sbintuitions/JMTEB}}, an ongoing effort inspired by the English MTEB~\cite{mteb}. For any given product query, the model must attempt to retrieve the description of relevant products among 149,999 product descriptions. The  main evaluation metric for this task is NDCG@10.

We provide an overview of key information on each dataset in table~\ref{tab:evaldatasets}. Despite the relative sparsity of Japanese retrieval benchmarks in comparison with English, we believe that these five datasets in these settings provide a good overview of retrieval models' capabilities on a wide array of real-world relevant usages.

\subsubsection{Development Evaluations}
\label{sec:devdata}

A large part of our study focuses on systematically evaluating a large variety of improvements to the ColBERT training and inference routine. As a result, we need a representative development set that is computationally inexpensive to run, while providing us with enough information to make decisions. We decide to use two evaluation sets, and report two key metrics for each of them. The first one is \textbf{JQaRA}, as presented above. We choose JQaRA due to its small size, being presented as a reranking task, while it has consistently shown a good ability to discriminate between models and a good correlation to performance on other datasets~\cite{jqara}. We report both NDCG@10 and MRR@10 as our development metrics.

As our second task, we follow ongoing efforts in creating lighter embeddings benchmarks\footnote{The team behind the English MTEB is currently developing a multilingual version of it, with a lite variant, including datasets similar to MIRACL-small, currently under discussion. \url{https://github.com/embeddings-benchmark/mteb/issues/784}} and introduce a smaller version of MIRACL's Japanese split, which we dub \textbf{MIRACL-small-ja}. This dataset is built through hard-negative mining. Using BM25, we retrieve the top 250 results for each of the 860 MIRACL development queries. We then enrich this data with all positive examples, if they were not present in the BM25 results. The resulting dataset contains 197,610 documents and 860 queries.

While the ideal situation would be for the final evaluation and development sets to be fully separate, this is not possible with the current amount of evaluation resources available. We pick these two datasets as we believe they provide a good balance between useful signal and minimising the risks of overfitting too strongly on these development evaluations, which would introduce a bias to our final results. 

\subsection{Dynamic Query Length}
\label{sec:querylen}

ColBERT models use a query augmentation mechanism, leveraging the use of \textsc{[MASK]} tokens, which are appended to every query, replacing traditional padding~\cite{colbert}, until the predefined maximum query length is reached. These tokens have been shown to learn different types of information, occasionally acting as term-importance weights~\cite{mask_analysis,mask2}. 

However, the exact impact of mask tokens and the best way to use them has been understudied. Instead of variable-length augmentation, a past study has explored simply appending 8 \textsc{[MASK]} tokens to every query, regardless of the actual query length~\cite{marginmse}. Another previous study has chosen to remove this augmentation mechanism entirely~\cite{colberter}. However, no study has compared the effects of these various choices against the original implementation.

We believe all three of these approaches to be suboptimal, and propose \textbf{dynamic query length} as a replacement. Dynamic query length effectively aims to improve the initial approach of padding each query with \textsc{[MASK]} tokens until the query maximum length by allowing it to more easily adapt to longer queries, while also borrowing from fixed-length augmentation for edge cases.

Effectively, our approach is to set the maximum query length to the nearest higher multiple of 32 (ColBERT's original maximum query length) before performing the \textsc{[MASK]}-padding. Additionally, if fewer than 8 augmentation tokens would be appended with the new query length, we ensure that at least 8 tokens are appended, overriding the maximum length.

To select the default query augmentation mechanism to use for JaColBERTv2, we evaluate all four of the discussed approaches.

\subsubsection{Results}

\begin{table}[H]
\begin{tabular}{l|cc|cc|cc}
                            & \multicolumn{2}{c|}{JQaRA} & \multicolumn{2}{c|}{MIRACL-small-ja} & \multicolumn{2}{c}{Average}     \\
                            & NDCG@10          & MRR@10  & NDCG@10              & Recall@5      & NDCG@10        & Overall        \\ \hline
Baseline                    & 0.578            & \textbf{0.820}   & \textbf{0.681}       & \textbf{0.707}         & 0.619          & 0.691          \\
No Augmentation & 0.557            & 0.817   & 0.632                & 0.661         & 0.595          & 0.667          \\
Fixed 8 tokens aug. & 0.577            & 0.813   & 0.670                & 0.692         & 0.624          & 0.688          \\
Dynamic query length        & \textbf{0.581}   & \textbf{0.820}   & \textbf{0.681}       & \textbf{0.707}         & \textbf{0.631} & \textbf{0.700}
\end{tabular}
\caption{Results of various query augmentation methods on our development set.}
\label{tab:queryaug}
\end{table}

Results for this experiment are presented in Table~\ref{tab:queryaug}. The results are pretty clear. On all datasets, disabling [MASK] augmentation is consistently largely outperformed by all query augmentation approaches. Fixed 8-token query augmentation, while performing considerably better than no augmentation, is similarly outperformed by both fixed and dynamic query lengths on every dataset.

On JQaRA, where the query length fluctuates more and some queries are considerably longer than others, dynamic query length outperforms a flat, higher token limit, suggesting that appending too many [MASK] tokens can produce a slightly detrimental effect. An empirical analysis of the results obtained on JQaRA also reveals that dynamic query length has virtually no impact on queries that are noticeably shorter than the maximum query length in a fixed query length setting. However, it noticeably improves NDCG@10 on queries which are nearer the maximum length, and would thus lose most of the query augmentation mechanism. This explains the small increase in overall NDCG@10 on the full dataset.

On MIRACL, where all queries have similar token counts and are all well under 32 tokens, the results for fixed and dynamic query lengths are identical, as would be expected.

Based on the results of this experiment, every result we report on subsequent ablation, as well as the final model results, will be using dynamic query length.

\subsection{Training Setting}
\label{sec:trainsettings}

In this section, we will evaluate the impact of certain changes to common components of the retrieval model training pipeline. We will first explore concerns around the impact of using in-batch negatives in Section~\ref{sec:ibneg}. We next explore the optimal way of scheduling the model's training, and the relevance of the recently recently schedule-free training method~\cite{schedulefree} in Section~\ref{sec:scheduling}. We will then study the benefits of score normalization, applied to both teacher and student models, in Section~\ref{sec:norm}, as well as explore the use and impact of different commonly used knowledge distillation loss functions in Section~\ref{sec:loss}. 

Finally, we will present the results of all of our experimental small-scale runs and discuss their implications in Section~\ref{sec:trainresults}.

\subsubsection{In-Batch Negatives}
\label{sec:ibneg}

In-batch negatives (\textit{IBNeg}) are frequently used as a way to augment the training of retrieval models~\cite{ibneg}. Effectively, within a given training batch, the IBNeg approach treats every other query's positive documents as additional negative examples in a binary relevance classification exercise. The query's original positive example is treated as a positive label, and every other query's positive example becomes a negative example. The model is asked to predict relevance over those newly created [query, document] pairs, and a cross-entropy loss is calculated over this prediction and then added to the model's main training loss.

This method has shown modest but consistent across studies performance improvements when used with single-vector retriever models~\cite{ibneg,e5}. This method is added to the ColBERT training recipe in the paper introducing ColBERTv2, following the promising results obtained by other models~\cite{colbertv2}. This choice was not thoroughly evaluated, and its impact is therefore unknown.

However, we empirically note that the resulting cross-entropy loss values are two orders of magnitude lower than the ones from the main KL-Divergence loss used in typical ColBERTv2 training. Moreover, we hypothesize that in-batch negatives are an unnecessary signal for training multi-vector models using 32-way triplets for multiple reasons, especially in non-English, data-constrained settings. Firstly, the information obtained from distilling the ranking distribution of a strong cross-encoder model over 32 documents should carry a much stronger signal than binary relevance labelling between a positive document and randomly sampled negatives. Secondly, this loss relies on the positive examples consistently being well-annotated and true positives, which is not guaranteed to be the case with lossy annotation processes, or even the partially-automated positive selection used in ColBERTv2~\cite{colbertv2}. 

To confirm the validity of our hypothesis, we will train two separate ablation models on the same data.

\subsubsection{Scheduling}
\label{sec:scheduling}

The learning rate scheduler used for the training of neural networks has been shown to have a potentially large impact on the performances of the resulting model~\cite{schedulerimpact}. There are are a few common schedulers yielding strong results, such as WSD (Warmup-Stable-Decay)~\cite{wsd}, which increases the learning rate steadily before plateauing for the majority of training and entering a decay phase, which should ideally be performed on higher quality data, or Linear-Decay, where the model's learning rate increases during a fixed number of warmup steps before linearly lowering until the final training step, among others. The original ColBERTv2, as well as JaColBERTv2~\cite{jacolbert}, used linear decay scheduling.

However, while tuned learning-rate scheduling consistently outperforms non-scheduled approached, it is not without constraints. Notably, for the best performance, it requires knowing the total number of steps in advance in all cases, or has constraints such having a higher quality data mix for schedulers relying on them for their annealing phase. Moreover, an optimal schedule for a large number of steps is not guaranteed to work as well for lower data quantities. These constraints are especially noticeable for retrieval models, which are expected to be put to use on a wide variety of downstream uses with varying data distributions, and therefore benefit immensely from being able to easily resume training without huge performance impact.

Recently, schedule-free learning has been proposed~\cite{schedulefree}. This new approach, while not yet thoroughly tested across all domains, has empirically shown very encouraging results on a large number of benchmarks. In practice, it introduces additional calculations as part of the optimizer steps, allowing it to vary the learning rate without the need for a fixed, pre-defined schedule. This considerably simplifies both the pre-training and fine-tuning processes, as there is no need to optimise the scheduler used for training and similar parameters can be re-used for different data scales.

Schedule-free learning has been noted as potentially requiring a higher learning rate than scheduled learning, with optimal values empirically falling in the range of 1 to 10 times the original learning rate~\cite{schedulefree}. We thus conduct an experiment comparing the training setting used for JaColBERTv2 and its ablations and schedule-free learning, with learning rates set to 1x, 3x and 5x the original 0.0001 learning rate. For all experiments, we retain the AdamW~\cite{adamw} optimizer used in previous work.

\subsubsection{Score Normalisation}
\label{sec:norm}

In the current ColBERTV2 training recipe, scores are unnormalised: the raw logits outputted by the teacher cross-encoder are used as teacher scores, and the output of the maxsim scoring function is used as the student model's score. These scores are on different scales, as the theoretical range for maxsim score is [0, Number of Query Tokens], while the range for cross-encoder logits include negative numbers and is on a scale dependant on the original model's training, which varies teacher by teacher.

While some distillation losses can be seen as being partially robust to scale differences, which partially justified the use of KL-Divergence in the ColBERTv2 paper~\cite{colbertv2}, we believe the lack of normalisation to be suboptimal for two main reasons. The first one is that, given the large difference in scale, the loss calculation is likely to lead to a better approximation of information loss if operating on a similar scale. The second is that normalised scores allow the models to focus purely on the relative ranking of results, rather than absolute scores, the latter of which may provide less useful information due to the automated nature of triplet generations.

We experiment with two used normalization approaches: one where only the teacher scores are normalized~\cite{spladev3}, and one where both the teacher and student scores are normalized. In all cases, we use min-max normalization, defined as:

\begin{equation}
score_{normalized} = \frac{score - score_{least\_relevant}}{score_{most\_relevant} - score_{least\_relevant}}
\end{equation}

Effectively, this gives a score of 1 to the most relevant document identified by the teacher and 0 to the least relevant one, regardless of their absolute score. Every other score is then placed on this scale depending on their distance to those two scores.

\subsubsection{Loss functions}
\label{sec:loss}

For knowledge distillation in retrieval models, two loss functions are commonly used~\cite{splade,rocketqav2}: MarginMSE and KL-Divergence, the latter of which is the one used by ColBERTv2.

\textbf{MarginMSE}~\cite{marginmse} consists in computing the Mean-Squared Error on the difference in the \textit{margin} between the predictions of the model being trained and the teacher model. The margin defined is the difference between the score the model gives to the positive document and the score it gives to negative documents. In the case of n-way training, this margin is calculated over every [positive\_document\_score, negative\_document\_n\_score]. MarginMSE is thus computed as follows (where N is the batch size):
\begin{equation}
    \text{MarginMSE} = \frac{1}{N} \sum_{i=1}^{N} \max(0, margin - (\text{score}_{\text{teacher}}(x_i) - \text{score}_{\text{student}}(x_i)))^2
\end{equation}
Effectively, the training objective becomes for the student model's margin to reproduce the teacher's margin as closely as possible.

\textbf{KL-Divergence}~\cite{kld}, on the other hand, seeks to directly minimise the difference between the distributions of scores of the model being trained and its teacher. KL-Divergence loss is computed as follows (where N is the batch size):
\begin{equation}
    \text{KL-Div} = \frac{1}{N} \sum_{i=1}^{N} \sum_{\text{score}} P_{\text{teacher}}(\text{score}|x_i) \log \left(\frac{P_{\text{teacher}}(\text{score}|x_i)}{P_{\text{student}}(\text{score}|x_i)}\right)
\end{equation}
In effect, it computes an estimation of how much information appears to be lost between the teacher model's distribution and the student model's one, and minimising this loss becomes the primary training objective.

The use of either MarginMSE or KL-Divergence has been reported for knowledge distillation into various retrieval models. While both loss functions have been shown to be strictly superior to more traditional MSE-based losses~\cite{gravedistill,marginmse}, there has been little head-to-head comparison of the two. Recently, the SPLADE-V3 authors anecdotally reported noticing overall similar performances between the losses, with MarginMSE favouring recall and KL-Divergence precision, and opted for a mixed-loss approach for their final model, using a conjunction of both with a lower weight attributed to MarginMSE~\cite{spladev3}. However, SPLADE-V3 was trained on only 8-way triplets, considerably fewer than our 32-way approach.

Our hypothesis is that, given our training setting with numerous negatives and its reduced reliance on having a strong positive, as only the distribution of scores matter, KL-Divergence remains the optimal choice to train ColBERT models with knowledge distillation. In order to test this hypothesis, we will compare its results with MarginMSE, as well as well as mixes of both with various MarginMSE weighting: $\lambda=\{0.2, 0.1, 0.05\}$.

\subsubsection{Ablation Results}
\label{sec:trainresults}

We present the results of all the ablation runs related to experiments detailed in the previous subsections in Table~\ref{tab:trainingablationresults}. As we run our experiments sequentially, the results are presented in the order of experiments, and the baseline for each new category is the best performing approach of the previous one.

\begin{table}[H]
\centering
\begin{tabular}{l|cc|cc|cc}
                                      & \multicolumn{2}{c|}{JQaRA}                                & \multicolumn{2}{c|}{MIRACL-small-ja}                        & \multicolumn{2}{c}{Average}                               \\
                                      & \multicolumn{1}{l}{NDCG@10} & \multicolumn{1}{l|}{MRR@10} & \multicolumn{1}{l}{NDCG@10} & \multicolumn{1}{l|}{Recall@5} & \multicolumn{1}{l}{NDCG@10} & \multicolumn{1}{l}{Overall} \\ \hline
\textbf{In-Batch Negs. (IBNeg)}       &                             &                             &                             &                               &                             &                             \\
With IBNeg (baseline)                 & \textbf{0.581}              & \textbf{0820}               & 0.681                       & 0.707                         & \textbf{0.631}              & 0.697                       \\
Without IBNeg                         & 0.580                       & \textbf{0.820}              & \textbf{0.682}              & \textbf{0.713}                & \textbf{0.631}              & \textbf{0.699}              \\ \hline
\textbf{Scheduling}                   &                             & \textit{}                   & \textit{\textbf{}}          & \textit{\textbf{}}            & \textit{}                   & \textit{\textbf{}}          \\
Linear Decay (baseline)               & 0.581                       & 0.820                       & 0.681                       & 0.713                         & 0.631                       & 0.699                       \\
Schedule-free (1x LR, 1e-05)          & 0.576                       & 0.820                       & 0.669                       & 0.707                         & 0.623                       & 0.693                       \\
Schedule-free (3x LR, 3e-05)          & \textbf{0.581}              & \textbf{0.821}              & \textbf{0.683}              & \textbf{0.717}                & \textbf{0.632}              & \textbf{0.701}              \\
Schedule-Free (5x LR, 5-05)           & 0.575                       & 0.815                       & 0.681                       & 0.709                         & 0.628                       & 0.695                       \\ \hline
\textbf{Normalization}                &                             &                             &                             &                               &                             &                             \\
No norm (baseline)                    & 0.581                       & 0.820                       & 0.681                       & 0.713                         & 0.631                       & 0.699                       \\
Normalize teacher scores              & 0.565                       & 0.802                       & 0.680                       & \textbf{0.717}                & 0.623                       & 0.691                       \\
Normalize student \& teacher          & \textbf{0.585}              & \textbf{0.827}              & \textbf{0.691}              & 0.716                         & \textbf{0.638}              & \textbf{0.705}              \\ \hline
\textbf{Loss Function}                &                             &                             &                             &                               &                             &                             \\
KL-Divergence (baseline)              & \textbf{0.585}              & \textbf{0.827}              & \textbf{0.691}              & \textbf{0.716}                & \textbf{0.638}              & \textbf{0.705}              \\
MarginMSE                             & 0.583                       & \textbf{0.827}              & 0.672                       & 0.699                         & 0.628                       & 0.695                       \\
\textbf{Mixed (KL-Div $\lambda=1.0$)} &                             &                             &                             &                               &                             &                             \\
+ MarginMSE $\lambda=0.2$             & 0.576                       & 0.813                       & 0.688                       & \textbf{0.716}                & 0.632                       & 0.698                       \\
+ MarginMSE $\lambda=0.1$             & 0.582                       & 0.816                       & 0.687                       & 0.714                         & 0.635                       & 0.700                         \\
+ MarginMSE $\lambda=0.05$            & 0.578                       & 0.812                       & 0.691                       & 0.691                         & 0.635                       & 0.693                      
\end{tabular}
\caption{Results of all training settings ablation results, compared to the relevant baseline}
\label{tab:trainingablationresults}
\end{table}

\textbf{In-batch negatives} Our results on in-batch negatives confirm our hypothesis: they do not appear to provide useful training to the model, even resulting in slightly decreased performance on MIRACL-small-ja. We believe this to be due to the factors we have highlighted with the main one being that the signal from distilling a teacher's score distribution over 32 documents appears to constitute a strong enough learning signal on its own. Additionally, some of our positive examples may be false positives, and many of our negative examples are false negatives. In-batch negatives are costly to compute, as they require an additional scoring stage for every query against all the other queries' positive documents, and materialising and computing a cross-entropy matrix. Moreover, they place additional constraints on training as we need to ensure higher data quality to fully leverage them. Removing the use of in-batch negatives thus represents an efficiency gain by lowering the compute and memory use of training at no performance cost. Following the results of this experiment, \textbf{we remove the use of in-batch negatives from our training recipe}.

\textbf{Scheduling} The results of our experiments on learning-rate schedulers highlight schedule-free learning~\cite{schedulefree} as a strong alternative to the Linear-Decay scheduler used in standard ColBERTv2 training. With a slight tweak to the learning rate, increasing it from the best-performing linear decay rate of $1e-05$ to $3e-05$, schedule-free learning results in a slight performance increase across the board. Additionally, schedule-free learning drastically reduces the level of optimisations required to operate at different data scales, and removes any costs associated with continuously stopping and restarting our model's training to expose it to different data distributions when attempting to use it in a different domain. \textbf{We choose to use schedule-free learning as part of our training}. This leads to us disabling the use of gradient clipping~\cite{gradientclipping}, which has been observed to cause schedule-free learning runs to fail to converge~\cite{schedulefree}.

\textbf{Normalization} Normalizing both teacher and student scores results in a sizeable performance increase on all datasets, while normalizing only teacher scores yield a consistent performance decrease. This is in line with our expectations. Moreover, normalized teacher scores, which appear to only function well when used in conjunction with normalized student scores, are a prerequisite to being able to best utilise an ensemble of teachers outputting scores on different scales. \textbf{Based on these clear results and this clear constraint, we introduce teacher and student score normalization to our training method}.

\textbf{Loss Functions} Unlike our other experiments, our empirical results demonstrate that the currently most commonly used loss function for knowledge distillation in ColBERT models, KL-Divergence, is the best performing option. MarginMSE exhibits reduced performance on all datasets, with it being noticeably more pronounced on MIRACL-small-ja. Interestingly, combining MarginMSE and KL-Divergence consistently results in worse results than using either loss on its own. We hypothesize that the varying quality of positive examples in our dataset could be a partial explanation for the substandard performance of MarginMSE, as it is calculated based on the margin between the positive example and negative examples, rather than strictly focusing on the score distribution in the same way as KL-Divergence. As it performs strictly worse than KL-Divergence and additionally frees us for data quality constraints,\textbf{we choose to discard MarginMSE and retain the KL-Divergence loss function for our model training}.

\subsection{Teacher Models}
\label{sec:teacher}

Another important part of knowledge distillation is the choice of teacher model. In Information Retrieval, teacher scores are most often obtained from the logits of cross-encoder reranker models, which assign a relevance score to query-document pairs~\cite{crossencdistil}. Cross-encoders are powerful retrieval models, as they are aware of both the query and the document at scoring time, whereas most other retrieval methods encode documents and queries separately~\cite{crossenc}. However, this means that they are particularly costly, as the model must run a full forward pass on every single pair in order to be able to output a score. As a result, they're unsuitable as a single retriever for medium-to-large scale document collections, but are particularly powerful as teachers for distillation.

The original ColBERTv2 paper used scores from a lightweight 22 million parameter MiniLM~\cite{minilm}-based model, itself distilled from a larger, more powerful cross-encoder~\cite{beir}. This choice was partially motivated due to the computational requirements of generating teacher scores for the entirety of the ColBERTv2 training set. As it comprises of over 19 million triplets, each made up of 64 individual query-document pairs, the score generation process for ColBERTv2 required computing cross-encoder scores for a total of 1.2 billion pairs.

However, larger cross-encoders generally result in better performance~\cite{beir,monot5}, and successful distilled models in the general domains have shown that stronger teachers yield stronger distilled models~\cite{minilm,distilbert}. Specifically, anecdotal results have shown that using T5-3B based rerankers such as MonoT5-3B~\cite{monot5}, based on a 3 billion parameters sequence-to-sequence model~\cite{t5}, consistently resulted in noticeably stronger distilled multi-vector models than distillation from smaller rerankers~\cite{xcolbert}.

\subsubsection{Single-Teacher}

\textbf{Models} To investigate the performance of various rerankers as teachers on Japanese retrieval, we generate teacher scores using a wide variety of models. As for monolingual Japanese models, we use the
\textsc{jp-small}\footnote{\url{https://huggingface.co/hotchpotch/japanese-reranker-cross-encoder-small-v1}} reranker, based on Multilingual-MiniLM~\cite{minilm}, the
\textsc{jp-base}\footnote{\url{https://huggingface.co/hotchpotch/japanese-reranker-cross-encoder-base-v1}} and
\textsc{jp-large}\footnote{\url{https://huggingface.co/hotchpotch/japanese-reranker-cross-encoder-large-v1}} rankers, based on Nagoya University's SIMCSE models~\cite{simcse}.
We also generate scores using the BGE-M3-reranker~\cite{m3} (\textit{M3}) model, the highest performing multi-lingual rerankers, as well as BGE-M3-jp\footnote{\url{https://huggingface.co/hotchpotch/japanese-bge-reranker-v2-m3-v1}} (\textit{M3-jp}), a version of it further fine-tuned on small-scale Japanese datasets. We selected these models among the available Japanese rerankers for their strong results on existing benchmarks~\cite{tateno}.
Finally, leveraging the fact that MMarco is a translation of MS Marco, we report the baseline performance of using the original ColBERTv2 triplets (\textit{original}), re-used in JaColBERTv2. Finally, we evaluate using the scores of the MonoT5-3B model mentioned above, also generated on the English version of the dataset.

\begin{table}[H]
\centering
\begin{tabular}{l|cc|cc|cc}
                        & \multicolumn{2}{c|}{JQaRA}                                & \multicolumn{2}{c|}{MIRACL-small-ja}                        & \multicolumn{2}{c}{Average}                               \\
                        & \multicolumn{1}{l}{NDCG@10} & \multicolumn{1}{l|}{MRR@10} & \multicolumn{1}{l}{NDCG@10} & \multicolumn{1}{l|}{Recall@5} & \multicolumn{1}{l}{NDCG@10} & \multicolumn{1}{l}{Overall} \\ \hline
Baseline                & 0.585                       & 0.827                       & 0.691                       & 0.716                         & 0.638                       & 0.705                       \\ \hline
\textbf{Single Teacher} &                             &                             &                             &                               &                             &                             \\
jp-small                & 0.567                       & 0.807                       & 0.703                       & 0.737                         & 0.635                       & 0.704                       \\
jp-base                 & 0.569                       & 0.81                        & 0.713                       & 0.747                         & 0.641                       & 0.701                        \\
jp-large                & 0.577                       & 0.816                       & 0.713                       & 0.741                         & 0.645                       & 0.712                       \\
M3                      & \textbf{0.589}              & 0.836                       & \textbf{0.740}              & \textbf{0.788}                & \textbf{0.665}              & \textbf{0.738}              \\
M3-jp                   & 0.588                       & \textbf{0.838}              & 0.728                       & 0.757                         & 0.658                       & 0.728                       \\
MonoT5-3B               & 0.587                       & 0.835                       & 0.594                       & 0.642                         & 0.591                       & 0.665                      
\end{tabular}
\caption{Results of ablation runs using various models as distillation teachers}
\label{tab:singleteacherresults}
\end{table}

\textbf{Results} The results of models trained using various models to generate teacher scores are presented in Table~\ref{tab:singleteacherresults}. The impact of using different teachers is immediately clear. Interestingly, the \textit{jp-*} models, in \textit{small}, \textit{base} and \textit{large} sizes, all yield stronger results on MIRACL-small-ja than the original teacher's scores. However, their use as teachers for our models appear to result in a significant performance decrease on JQaRA compared the original scores. On the other hand, the BGE-Reranker-M3 models prove to be very strong teachers. Interestingly, the multi-lingual version of M3, which has not been fine-tuned on Japanese data, vastly outperforms its fine-tuned version on MIRACL-small-ja, and roughly matches its performance on JQaRA, resulting in a noticeably stronger average performance.  Ultimately, BGE-Reranker-M3, in its non-finetuned multilingual version, appears to be strongest available teacher, reaching an overall score of 0.738, compared to the noticeably lower 0.705 score of the original training scores.

Finally, it is also worth noting that the MonoT5-3B reranker yields strong results on JQaRA, but leads to a sizeable performance drop on MIRACL-small-ja. This behaviour indicates a potentially interesting trend: models generating scores on English-language MSMarco rather than MMarco, as is the case for the original scores and MonoT5-3B, appears to result in strong performance on JQaRA and weaker results on MIRACL-small-ja, while the opposite holds true for Japanese-language models.\footnote{ We do not explore this phenomenon further as generating scores with a 3 billion parameter model is particularly costly, and this analysis is outside the scope of this study.}.

\subsubsection{Ensembled Teachers}

\textbf{Ensembling Teachers} Additionally, recent work on English Information Retrieval has increasingly highlighted that ensembling the scores of multiple teachers produce better distilled models~\cite{spladev3}, even when the ensembled teachers' individual performance are largely similar~\cite{marginmse}. The most common way of ensembling multiple teachers' scores is via a two-step process, where the scores are first normalized using min-max normalization as in Section~\ref{sec:norm}, before being averaged. We believe that these results may not reproduce in a Japanese setting, as there exists far fewer Japanese base models and strong rerankers than in high-resource languages settings. However, we evaluate a large range of teacher ensembling as part of our study, using the teacher models described above.

\begin{table}[H]
\centering
\begin{tabular}{l|cc|cc|cc}
                            & \multicolumn{2}{c|}{JQaRA}                                & \multicolumn{2}{c|}{MIRACL-small-ja}                        & \multicolumn{2}{c}{Average}                               \\
                            & \multicolumn{1}{l}{NDCG@10} & \multicolumn{1}{l|}{MRR@10} & \multicolumn{1}{l}{NDCG@10} & \multicolumn{1}{l|}{Recall@5} & \multicolumn{1}{l}{NDCG@10} & \multicolumn{1}{l}{Overall} \\ \hline
Baseline (orig.)            & 0.585                       & 0.827                       & 0.691                       & 0.716                         & 0.638                       & 0.705                       \\
Best single teacher (M3)    & 0.589                       & 0.836                       & \textbf{0.740}              & \textbf{0.788}                & 0.665                       & \textbf{0.738}              \\ \hline
\textbf{Ensembled Teachers} & \multicolumn{1}{l}{}        & \multicolumn{1}{l|}{}       & \multicolumn{1}{l}{}        & \multicolumn{1}{l|}{}         & \multicolumn{1}{l}{}        & \multicolumn{1}{l}{}        \\
M3 + M3-jp                  & 0.587                       & 0.835                       & 0.733                       & 0.764                         & 0.660                       & 0.730                       \\
M3 + M3-jp + orig.          & 0.585                       & 0.832                       & 0.742                       & \textit{0.778}                & 0.664                       & 0.734                       \\
M3 + M3-jp + large          & 0.587                       & 0.828                       & 0.739                       & 0.766                         & 0.663                       & 0.730                       \\
M3 + M3-jp + large + base   & 0.589                       & 0.834                       & 0.738                       & 0.763                         & 0.664                       & 0.731                       \\
M3 + MT5                    & \textit{\textbf{0.598}}     & \textit{\textbf{0.837}}     & 0.702                       & 0.745                         & 0.650                        & 0.721                       \\
M3-jp + MT5                 & 0.596                       & 0.832                       & 0.694                       & 0.734                         & 0.645                       & 0.714                       \\
M3 + M3-jp + MT5            & \textit{\textbf{0.598}}     & 0.832                       & \textit{\textbf{0.743}}     & 0.769                         & \textbf{0.671}              & \textit{0.736}              \\
M3 + orig. + MT5            & 0.597                       & 0.837                       & 0.713                       & 0.747                         & 0.655                       & 0.724                       \\
M3-jp + orig. + MT5         & 0.596                       & 0.835                       & 0.707                       & 0.740                         & 0.652                       & 0.720                       \\
M3 + M3-jp + orig. + MT5    & 0.596                       & 0.835                       & 0.727                       & 0.760                         & 0.662                       & 0.730                       \\
All rerankers               & 0.589                       & \textit{0.841}              & 0.711                       & 0.759                         & 0.65                        & 0.725                      
\end{tabular}
\caption{Results of ablation runs using various ensembles of models as distillation teachers. Best overall results are reported in bold, and best results within the ensembled category in italic. \textit{"orig." refers to the original training set, "MT5" to to MonoT5, "large" to jp-large and "base" to jp-base.}}
\label{tab:ensembledteacherresults}
\end{table}

\textbf{Results} The results of various ensembling combinations are presented in Table~\ref{tab:ensembledteacherresults}. Overall, no ensembling combination outperforms simply using BGE-Reranker-M3 as a single teacher on all metrics, which confirms our initial impression that its individual performance outweighs any gains from ensembling with weaker models.

It is worth noting that some ensembling combinations, such as BGE-Reranker-M3+M3-jp+MonoT5-3B reach promising results, with higher NDCG@10 results on both development sets. However, generating teacher scores is costly, particularly with the very large MonoT5-3B model, and the weaker performance of the ensembled teachers on non-NDCG metrics does not appear to justify this computational cost. We thus choose to retain BGE-Reranker-M3 as our single-model teacher for the full training run.

\subsection{Post-Training}
\label{sec:posttrain}

As discussed in Section~\ref{sec:trainingdata}, our models will be trained on MMarco, a dataset which was machine translated from English to Japanese in 2019, and therefore often contains lower quality Japanese or odd sentence constructions.

While we believe that this issue is unlikely to have a large impact on our final model, we propose post-training (also called \textit{fine-tuning}) the model on smaller, higher quality datasets. To do so, we choose to use the previously discussed datasets MIRACL~\cite{miracl} and JQaRA~\cite{jqara}, as well as JaGovFaqs, a subcomponent of JMTEB containing questions and answers from the Japanese Government's FAQ sections.

\begin{table}[H]
\begin{tabular}{l|ccc}
                     & Triplets & Proportion (w/o MMarco) & Proportion (with MMarco) \\ \hline
MIRACL               & 115,365  & 64.01\%                 & 60.72\%                  \\
JQaRA                & 35,733   & 19.93\%                 & 14.17\%                  \\
JaGovFaqs            & 28,902   & 16.06\%                 & 15.21\%                  \\
Total without MMarco & 180,000  & 100\%                   & 90.9\%                   \\ \hline
MMarco               & 18,000   & 10\% (for reference)    & 9.9\%                    \\
Total with MMarco    & 198,000  & 110\% (for reference)   & 100\%                   
\end{tabular}
\caption{Overview of the datasets used in the post-training data mixes.}
\label{tab:posttrain}
\end{table}

This post-training phase has two aims: the first is to improve our model's ability to understand Japanese in more diverse settings than the ones observed during pre-training. Our second aim is to highlight the potential gains that can be obtained with relatively small scale post-training on domain specific data, which should be reflected in potential performance improvements on JQaRA and MIRACL.

We present the full make-up of the post-training data in~\ref{tab:posttrain}. We experiment with two post-training settings. The first setting only contains the three datasets listed above. The second one additionally includes 10\% data randomly sampled from the MMarco triplets used for pre-training, to address the issue of catastrophic forgetting, where the model forgets previous training when exposed to exclusively new data~\cite{catastrophicforgetting}

\subsection{Checkpoint Averaging}
\label{sec:averaging}

Finally, we present the hypothesis that checkpoint averaging can improve our model's generalisation ability. Checkpoint averaging, also called "model merging", consists in taking multiplied different checkpoints of similarly sized models and averaging their parameters at each layer to create a merged model.

This practice has a long history in statistical Machine Learning~\cite{polyak}, with early research showing that averaging the parameters of a trained model during its learning rate decay phase can outperform the final checkpoint~\cite{polyak2}. More recently, this method has experienced renewed interest, with the merging of different Large Language Models (LLMs) showing noticeably stronger benchmark results than    a single checkpoint~\cite{sakana}. During the writing of this work, Meta released the Llama-3.1 family of models, where the final models consist of an averaged version of the final few checkpoints~\cite{llama3}, following the intuition of Polyak's method~\cite{polyak3}.

In Information Retrieval, this practice is largely understudied, with no previous work reporting its use to the best of our knowledge. We believe that it is especially suitable to creating better overall retrievers by merging the weights of post-trained (\textit{fine-tuned}) models, which might hurt performance on datasets out of its fine-tuning distribution, with the weights of the original model. We hypothesize that doing so will return most of the performance improvements on datasets similar to the post-training set while avoiding degradation on other tasks.

\section{Final Experimental Setting}
\label{sec:finalsettings}

The final experimental setting, used for our model training, is deduced from the results of the various ablation experiments detailed in the previous sections. We provide an overview of the ablation-informed decisions made in Table~\ref{tab:finalsetting}.

\begin{table}[H]
\centering
\begin{tabular}{l|cc}
\multicolumn{1}{c|}{\textbf{Setting}} & \textbf{Setting Used} & \textbf{Changed} \\ \hline
Query-Length                          & Dynamic               & Yes              \\
In-Batch Negatives                    & No                    & Yes              \\
Scheduler                             & Schedule-Free         & Yes              \\
Gradient Clipping                     & Disabled              & Yes              \\
Learning Rate                         & 3e-5                  & Yes              \\
Teacher Scores Normalization          & Min-Max Normalization & Yes              \\
Student Scores Normalisation          & Min-Max Normalization & Yes              \\
Mixed Loss                            & No                    & No               \\
Loss Function                         & KL-Divergence         & No               \\
Teacher Model                         & BGE-M3-Reranker       & Yes              \\
\textit{Batch size (per GPU)}         & \textit{16}           & \textit{No}               \\
\textit{Maximum Document Length}      & \textit{300}          & \textit{No}               \\
\textit{Warmup steps}                 & \textit{5\% of total} & \textit{No}               \\
\textit{Gradient Accumulation}        & \textit{Disabled}     & \textit{No}              
\end{tabular}
\caption{Overview of the final optimal training settings resulting from our experiments, and whether or not they represent a change from previously used settings. Settings in \textit{italic} are retained from previous approaches with no further experiments within this paper.}
\label{tab:finalsetting}
\end{table}

We have shown that using dynamic query length at inference-time is strictly superior to fixed-length queries in Section~\ref{sec:querylen}, and thus adopt dynamic query length for all our final evaluations.

As for our training setting, following the ablation results, we have confirmed the hypothesis presented in Section~\ref{sec:ibneg}, and opt not use in-batch negatives. We adopt schedule-free learning, as presented in Section~\ref{sec:scheduling}, due to its reduced constraints with no performance decrease. As suggested in Section~\ref{sec:norm}, we normalize both teacher and student scores using min-max normalization, and use KL-Divergence loss, presented in Section~\ref{sec:loss}. We train our models using knowledge distilled from the teacher scores of BGE-M3, as our results in Section~\ref{sec:teacher} show it to be the best performing option on our development sets. We also report training settings that are unchanged from JaColBERTv2's original settings\footnote{After carefully analysing the data make-up of the ablation dataset and verifying it would not have a strong impact, maximum document length was set to 228 for ablation runs in order to minimise computational cost. For the full length training run, we reverted to JaColBERTv2's 300 to account for longer outlier documents.}, to facilitate reproduction.\\

\textbf{Potential Impact on Data Selection} An important aspect of our training recipe changes are that the labels of "positives" and "negatives" for each document become unused. Indeed, our only learning metric is based on the KL-Divergence loss on min-max normalized teacher scores, which trains our model to attempt to learn the score distribution of its teacher model. While we do not further modify the training data mix in this work, this lifts a considerable constraint for future endeavours, as the use of positive and negative labels often renders the curation of training data more difficult due to the porous nature of relevance judgements. Indeed, curating the proper mix of "hard" and "easy" negatives is a complex process whose outcomes are not yet fully understood~\cite{waterloohardnegs}, and some "hard negatives" could very well be positives, and filtering "too-hard-negatives" remains a mostly empirical, error-prone process~\cite{snowflake}.

\textbf{Comparison to previous models} Table~\ref{tab:finalsettingsresults} presents a comparison of an ablation run with the original JaColBERTv2's training settings, an ablation run comprising of all our final's chosen parameters, JaColBERTv1, our base model, and JaColBERTv2, the previous best-performing Japanese ColBERT model. Our results appear to support our claim that the original JaColBERTv2 training recipe is highly suboptimal. With just 320,000 32-way triplets, our ablation model vastly outperforms the original setting, reaching an average score of 0.738 while the original settings only marginally outperforms JaColBERTv1, with a score of 0.675. Even more noteworthy, our ablation run, with just 4\% of JaColBERTv2's training data, outperforms it on both development evaluation sets. Finally, it is important to note that all models discussed here outperform JaColBERTv1's score of 0.674, further showcasing the importance of nway-training and knowledge distillation, even in sub-optimal training settings.

\begin{table}[H]
\centering
\begin{tabular}{l|cc|cc|cc}
                                                                   & \multicolumn{2}{c|}{JQaRA}                                & \multicolumn{2}{c|}{MIRACL-small-ja}                        & \multicolumn{2}{c}{Average}                               \\
                                                                   & \multicolumn{1}{l}{NDCG@10} & \multicolumn{1}{l|}{MRR@10} & \multicolumn{1}{l}{NDCG@10} & \multicolumn{1}{l|}{Recall@5} & \multicolumn{1}{l}{NDCG@10} & \multicolumn{1}{l}{Overall} \\ \hline
\begin{tabular}[c]{@{}l@{}}JaColBERTv1\\ (Base Model)\end{tabular} & 0.550                       & 0.811                       & 0.652                       & 0.681                         & 0.601                       & 0.674                       \\
JaColBERTv2                                                        & 0.585                       & \textbf{0.836}              & 0.727                       & 0.751                         & 0.656                       & 0.725                       \\
Original settings                                                  & 0.578                       & 0.820                       & 0.681                       & 0.707                         & 0.630                       & 0.697                       \\
Final Settings                                                     & \textbf{0.589}              & \textbf{0.836}              & \textbf{0.740}              & \textbf{0.788}                & \textbf{0.665}              & \textbf{0.738}             
\end{tabular}
\caption{Results of JaColBERTv1 (the base model for both our experiments and JaColBERTv2), JaColBERTv2, the original training settings as well as our final experimental setting ablation runs on our development evaluation set. Best results are indicated in \textbf{bold}.}
\label{tab:finalsettingsresults}
\end{table}

\textbf{Hardware Usage} As part of these experiments, a total of 28 ablation models were trained. Each training run represents 1.3 hours of training time on 4 A100 GPUs, or 5.2 A100 hours per run. As a result, we estimate that a total of 146 A100 hours were spent on ablation runs. Additionally, 15 A100 hours were spent on generating teacher scores over the full 3,200,000 training set using the BGE-M3-Reranker model. Finally, an estimated 8 GPU hours were spent generating teacher scores with all models for the ablation runs, the majority of it dedicated to the MonoT5-3B model. This results in a total pre-final training GPU usage of \textbf{169 A100 hours}.

\section{Final Results}

The results for all newly introduced models and the baselines are presented in Table~\ref{tab:finalresults}. In the interest of readability, we report the results for all models on the main evaluation of each dataset, as defined in Section~\ref{sec:evaldata}. Full evaluation results of our new models covering other, less commonly reported metrics, are available in Appendix~\ref{app:fullresults}.

\begin{table}[H]
\centering
\begin{tabular}{lccccc|c}
\multicolumn{1}{c}{}                                                                   & \begin{tabular}[c]{@{}c@{}}JSQuAD\\ Recall@3\end{tabular} & \begin{tabular}[c]{@{}c@{}}MIRACL\\ NDCG@10\end{tabular} & \begin{tabular}[c]{@{}c@{}}JQaRA\\ NDCG@10\end{tabular} & \begin{tabular}[c]{@{}c@{}}JaCWIR\\ MAP@10\end{tabular} & \begin{tabular}[c]{@{}c@{}}ESCI\\ NDCG@10\end{tabular} & Average        \\ \hline
\textbf{Baselines (Multi)}                                                             &                                                           &                                                          &                                                         &                                                         &                                                        &                \\
bge-m3                                                                                 & 0.939                                                     & \textit{0.728}                                           & 0.539                                                   & 0.864                                                   & 0.399                                                  & 0.694          \\
bge-m3 (all)                                                                           & 0.958                                                     & \textit{0.752}                                                    & 0.576                                                   & 0.906                                                   & 0.380                                                  & 0.714          \\
multilingual-e5 (large)                                                                & 0.953                                                     & \textit{0.706}                                           & 0.554                                                   & 0.876                                                   & 0.320                                                  & 0.682          \\
multilingual-e5 (base)                                                                 & 0.934                                                     & \textit{0.647}                                           & 0.471                                                   & 0.852                                                   & 0.347                                                  & 0.650          \\
multilingual-e5 (small)                                                                & 0.934                                                     & \textit{0.636}                                           & 0.492                                                   & 0.869                                                   & 0.331                                                  & 0.652          \\ \hline
\textbf{Baselines (Mono)}                                                              &                                                           &                                                          &                                                         &                                                         &                                                        &                \\
GLuCoSE                                                                                & \textit{0.798}                                            & \textit{0.348}                                           & 0.309                                                   & 0.686                                                   & 0.207                                                  & 0.470          \\
sup-simcse-ja (base)                                                                   & 0.793                                                     & 0.171                                                    & 0.312                                                   & 0.578                                                   & 0.140                                                  & 0.399          \\
sup-simcse-ja (large)                                                                  & 0.777                                                     & 0.199                                                    & 0.392                                                   & 0.474                                                   & 0.140                                                  & 0.396          \\
JaColBERTv1                                                                            & 0.961                                                     & 0.583                                                    & 0.550                                                   & 0.904                                                   & 0.418                                                  & 0.683          \\
JaColBERTv2                                                                            & 0.968                                                     & 0.667                                                    & 0.585                                                   & 0.919                                                   & 0.462                                                  & 0.720          \\ \hline
\textbf{JaColBERTv2.5}                                                                 &                                                           &                                                          &                                                         &                                                         &                                                        &                \\
Final checkpoint                                                                        & 0.973                                                     & 0.756                                                    & 0.601                                                   & 0.928                                                   & 0.462                                         & 0.744          \\
+ full post-train                                                                      & 0.970                                                     & \textit{\textbf{0.780}}                                  & \textit{0.608}                                          & 0.924                                                   & 0.451                                                  & 0.747          \\
+ post-train (no mmarco)                                                               & 0.972                                                     & \textit{0.772}                                           & \textit{0.613}                                          & 0.923                                                   & 0.452                                                  & 0.746          \\
\begin{tabular}[c]{@{}l@{}}JaColBERTv2.4\\ (no post train, averaged)\end{tabular}      & 0.973                                                     & 0.757                                                    & 0.601                                                   & \textbf{0.929}                                          & \textbf{0.463}                                                  & 0.745          \\
\begin{tabular}[c]{@{}l@{}}JaColBERTv2.5 (final)\\ (post-train, averaged)\end{tabular} & \textbf{0.974}                                            & \textit{0.778}                                           & \textit{\textbf{0.618}}                                 & 0.928                                                   & 0.462                                                  & \textbf{0.752}
\end{tabular}
\caption{Results for all baselines and newly introduced models on the main metric for all five evaluation datasets, as well as their averaged results. Best overall results are indicated in \textbf{bold}. Results in \textit{italics} indicate that the model was exposed to the task's training set.}
\label{tab:finalresults}
\end{table}

\textbf{Initial Results} Immediately, we notice that all versions of JaColBERTv2.5, no matter the post-training and checkpoint averaging setup, largely outperform JaColBERTv2 and all previous approaches on all five benchmarks. The final checkpoint resulting from the initial training phase reaches an average score of 0.744, with JaColBERTv2 previously reaching 0.720 and BGE-M3 in its \textit{all} setting, combining dense, sparse and multi-vector representations, 0.714. More interestingly, this checkpoint also achieves an entirely out-of-domain performance of 0.756 NDCG@10 on MIRACL, largely surpassing JaColBERTv2's out-of-domain 0.667 and outperforming BGE-M3 (all)'s score of 0.0752, despite the latter having been trained on MIRACL. These very strong results confirm our intuition that the existing and most commonly used training recipe for multi-vector models, used in training JaColBERT, was largely suboptimal, and that our proposed improvements result in substantially stronger downstream results. Moreover, this confirms our intuition that it is possible to reach state-of-the-art Japanese retrieval performance while using two orders of magnitude fewer data and compute than leading models. 

\textbf{Post-Training} We then explore the results of our post-training step. While they both result in a slight average score improvement, bringing the average model scores to 0.747 when post-training with MMarco data re-injected and 0.746 without. However, these results are obtained via large gains on datasets which are now in-domain, namely MIRACL and JQaRA, while causing moderate degradation on all three other datasets. Interestingly, while adding MMarco data to the post-training set results in less pronounced performance increases on JQaRA, it counter-intuitively further increases the model's performance on MIRACL, despite reducing its relative importance in the training set.

\textbf{Checkpoint Averaging} Finally, it is noticeable that averaging the two best-performing checkpoints of the initial training with the final one result in a slight overall performance increase, although it does not represent a substantial improvement. As a result of the slight increase, we release this model as \textbf{JaColBERTv2.4}. However, averaging the two post-trained checkpoints, with and without MMarco, with the three checkpoints from the original run results in greatly improved performance across the board. Indeed, this final version of the model, which is the version we name \textbf{JaColBERTv2.5}, largely outperforms all other variants, reaching an average score of 0.752. This model retains most of the performance gain of the most successful post-training runs on MIRACL, resulting in an increase in the NDCG@10 score on it from 0.757 to 0.778, just 0.002 points short of the best post-trained model (0.780). Even more interestingly, it is the strongest performing model of JQaRA, representing a 0.017NDCG@10 increase on the non post-trained model, but also a 0.005 increase on the post-trained versions.

These gains are achieved with little or no degradation on any of the other datasets, with averaging fully compensating for the degradation experienced in the pre-averaging post-trained models. This suggests that checkpoint averaging has a strong ability to revert catastrophic forgetting~\cite{catastrophicforgetting} in retrieval models, while retaining most of the domain-specific performance gained from post-training.\\

\textbf{Hardware Usage} Our final training run took 15.5 hours on 4 A100 GPUs, representing a total GPU usage of 62 hours. Post-training without MMarco took an hour on the same hardware setting, 1.2 hours with MMarco, representing a total usage of 8.8 hours over the two runs, which we round up to 9 for clearer reporting. Our final training and post-training steps, in total, required 70 A100 hours. Combined with the 162 GPU hours budget spent on generating teacher scores, the total GPU usage of this study represents \textbf{233 A100 hours}. While slightly above the JaColBERTv2 computational budget of 228 A100 hours, we fall within the upper bound of our allocated computational budget of 239 hours, while reaching significantly stronger results than any previous approach. 

\section{Conclusion}

In this work, we present JaColBERTv2.5, a model obtained by systematically evaluating the impact of potential improvements to the ColBERTv2 inference and training recipes, through small-scale ablation runs. Notably, we identify a better way to handle inference-time query length, devise a much improved training setting, and identify the optimal teacher model for knowledge distillation.

JaColBERTv2.5, while trained on only 40\% as much data as JaColBERTv2, largely outperforms all other retrieval methods in Japanese, including multilingual models with five times the parameter count and trained using two orders of magnitude more compute and data.

Throughout our experiments, we have shown that multi-vector retrieval models can not only effectively bridge the gap between multilingual models and monolingual, but largely outperform the former, with only moderate compute resources and lower quality data than their English equivalents.

The results of our training recipe also show that it is possible to train models, using knowledge distillation, without any reliance on hard "positive" or "negative" labels, focusing entirely on the teacher score distribution instead. This represents valuable information for future work, as a step towards greatly simplifying the data curation process.

We have also demonstrated that checkpoint averaging, where the weights of multiple checkpoints of a similarly-shaped model are averaged to create a \textit{merged model}, can greatly improve the generalisation potential of fine-tuned JaColBERT models, while retaining the same out-of-domain performance as the original model.

We make both JaColBERTv2.5\footnote{\url{https://huggingface.co/answerdotai/JaColBERTv2.5}}, our final checkpoint resulting from averaging our final model with post-training runs on two slightly different distributions, and JaColBERTv2.4\footnote{\url{https://huggingface.co/answerdotai/JaColBERTv2.4}}, the outcome of merging the three best checkpoints of the original pre-training, publicly available.

We believe that our work can support the development of future mono-lingual retrievers, both in Japanese and other lower-resources languages. Notably, we believe that our improved training recipe can be directly applied to sparse retrieval models such as SPLADE~\cite{spladev3}, which has already shown strong potential in a Japanese setting\footnote{As demonstrated by an early release from the University of Tsukuba available at \url{https://huggingface.co/aken12/splade-japanese-v3}.}

In order to best support such future work, we make the entirety of our training data for both ablation and full training runs, teacher scores included, publicly available\footnote{\url{https://huggingface.co/datasets/answerdotai/MMarco-japanese-32-scored-triplets}}. To further studies into better understanding of multi-vector retrieval models, we release all mid-training checkpoints, saved every 2,000 training steps\footnote{\url{https://huggingface.co/collections/bclavie/jacolbertv25-checkpoints-66a37d8da6b0d4d69c14f9c3}}.

Finally, while our specific application case is focused on the Japanese language, all of our training recipe improvements are language-agnostic, and even our ablation-sized models, trained on just 320,000 triplets, vastly outperform previous monolingual approaches. As a result, we believe that our method can be directly applied to other languages and domains and yield large performance gains.

\section{Ethical Considerations}

We acknowledge the importance of ethical consideration in Natural Language Processing work. We have ensured to make our work as reproducible as possible, making both the the final model, in-development versions of it, and the entirety of our training data publicly available to facilitate reproduction and future work.

There are no extreme ethical risks associated with our models. However, while they are not generative and will not, by themselves, generate harmful content, they fall in line with existing retrieval work. As such, our work is not exempt from potential biases, especially as the largest part of our training data is a lightly filtered internet corpus which then underwent machine translation. It is possible that our models might unduly favor certain types of content, and may rank misinformation or harmful content highly for certain queries. 

\section{Acknowledgement}

The author thank Yuichi Tateno for his extensive work in both creating benchmarks for, as well as training and benchmarking, Japanese rerankers, and Hayato Tsukagoshi for his willingness to share his work on Japanese SimCSE models and exploring the Japanese embedding training and data landscape. Further thanks extend to Benjamin Warner for sharing his expert advice on the various ways to optimise model training and detect inefficiencies, especially in the area of scheduling, as well as Alexis Gallagher, for very insightful feedback during the writing of this work. The author would also like acknowledge the helpful exchanges with Omar Khattab, Antoine Chaffin and Griffin Adams for their eagerness to discuss and help clarify ideas, as well as Professor Makoto P. Kato for helpful exchanges around building better Japanese retrieval models.

\begin{CJK*}{UTF8}{min}
\bibliographystyle{acm}
\bibliography{bibliography}
\end{CJK*}

\appendix

\section{Full JaColBERTv2.5 results}
\label{app:fullresults}

\begin{table}[H]
\begin{tabular}{lc|ccccc}
                & Baseline                                                                    & \multicolumn{5}{c}{Our Models}                                                                                                                                                                                                                                                                                    \\
                & \multicolumn{1}{l|}{\begin{tabular}[c]{@{}c@{}}JaColBERT\\ v2\end{tabular}} & \begin{tabular}[c]{@{}c@{}}JaColBERT\\ v2.5\end{tabular} & \begin{tabular}[c]{@{}c@{}}JaColBERT\\ v2.4\end{tabular} & \begin{tabular}[c]{@{}c@{}}Final\\ Checkpoint\end{tabular} & \begin{tabular}[c]{@{}c@{}}Post-Train\\ (no mmarco)\end{tabular} & \begin{tabular}[c]{@{}c@{}}Post-Train\\ (full)\end{tabular} \\ \hline
\textbf{MIRACL} &                                                                             &                                                          &                                                          &                                                            &                                                                  &                                                             \\
NDCG@10         & 0.667                                                                       & \textit{0.778}                                           & 0.757                                                    & 0.756                                                      & \textit{0.772}                                                   & \textit{\textbf{0.780}}                                     \\
MRR@10          & 0.688                                                                       & \textit{0.795}                                           & 0.780                                                    & 0.781                                                      & \textit{0.793}                                                   & \textit{\textbf{0.802}}                                     \\
Recall@10       & 0.802                                                                       & \textit{\textbf{0.887}}                                  & 0.869                                                    & 0.871                                                      & \textit{0.880}                                                   & \textit{\textbf{0.887}}                                     \\
Recall@100      & 0.961                                                                       & \textit{0.989}                                           & 0.987                                                    & 0.987                                                      & \textit{0.987}                                                   & \textit{\textbf{0.990}}                                     \\ \hline
\textbf{JQaRA}  &                                                                             &                                                          &                                                          &                                                            & \textit{}                                                        & \textit{}                                                   \\
NDCG@10         & 0.585                                                                       & \textit{\textbf{0.618}}                                  & 0.601                                                    & 0.601                                                      & \textit{0.613}                                                   & \textit{0.608}                                              \\
MRR@10          & 0.836                                                                       & \textit{\textbf{0.856}}                                  & 0.846                                                    & 0.843                                                      & \textit{0.849}                                                   & \textit{0.846}                                              \\ \hline
\textbf{JaCWIR} &                                                                             &                                                          &                                                          &                                                            &                                                                  &                                                             \\
MAP@10          & 0.919                                                                       & 0.928                                                    & \textbf{0.929}                                           & 0.928                                                      & 0.923                                                            & 0.924                                                       \\
Hit Rate@10     & \textbf{0.982}                                                              & 0.979                                                    & 0.980                                                    & 0.979                                                      & 0.979                                                            & 0.978                                                       \\ \hline
\textbf{JSQuAD} &                                                                             &                                                          &                                                          &                                                            &                                                                  &                                                             \\
Recall@1        & 0.917                                                                       & \textbf{0.930}                                           & \textbf{0.930}                                           & \textbf{0.930}                                             & 0.928                                                            & 0.925                                                       \\
Recall@3        & 0.967                                                                       & \textbf{0.974}                                           & \textbf{0.974}                                           & \textbf{0.974}                                             & 0.972                                                            & 0.970                                                       \\
Recall@5        & 0.976                                                                       & \textbf{0.982}                                           & \textbf{0.982}                                           & \textbf{0.982}                                             & 0.980                                                            & 0.978                                                       \\
Recall@10       & 0.982                                                                       & \textbf{0.987}                                           & \textbf{0.987}                                           & \textbf{0.987}                                             & 0.986                                                            & 0.986                                                       \\ \hline
\textbf{ESCI}   &                                                                             &                                                          &                                                          &                                                            &                                                                  &                                                             \\
NDCG@10         & 0.462                                                                       & 0.462                                                    & \textbf{0.463}                                           & 0.462                                                      & 0.452                                                            & 0.451                                                       \\
MRR@10          & 0.619                                                                       & 0.619                                                    & 0.621                                                    & \textbf{0.622}                                             & 0.610                                                            & 0.606                                                       \\
Recall@10       & 0.381                                                                       & \textbf{0.386}                                           & \textbf{0.386}                                           & \textbf{0.386}                                             & 0.379                                                            & 0.376                                                      
\end{tabular}
\caption{Presentation of the full results of different variants of our newly introduced models across a range of metrics. Best results are presented in \textbf{bold}. Results in \textit{italic} indicate that the task is in-domain.}
\label{tab:appendixfullresults}
\end{table}

\end{document}